\documentclass[12pt, a4paper]{article}
\usepackage{amssymb}
\usepackage{amsmath}
\usepackage{graphicx}
\usepackage{subfigure}
\usepackage{color}
\usepackage{multicol}
\usepackage[percent]{overpic}
\usepackage[T1]{fontenc}
\usepackage[utf8]{inputenc}
\usepackage{authblk}

\usepackage[round]{natbib}
\def\MC{\mathcal{C}}
\def\MP{\mathcal{P}}
\def\CD{\mathcal{D}}
\def\BP{\mathbb{P}}
\def\BE{\mathbb{E}}
\def\rd{{\rm{d}}}

\usepackage{amssymb}
\usepackage{gensymb}

\title{ABC for climate: dealing with expensive simulators}
 \author[1,*]{P. B. Holden}
 \author[1]{N. R. Edwards}
 \author[2]{J. Hensman} 
 \author[3]{R. D. Wilkinson}
 
 \affil[1]{Earth, Environment and Ecosystems, Open University, UK}
 \affil[2]{Computer Science, University of Sheffield, UK}
 \affil[3]{School of Mathematics and Statistics, University of Sheffield, UK}
\affil[*]{\it Corresponding author}

\begin{document}
\maketitle

 \section{Introduction}
 
One of the primary challenges faced when calibrating a simulator using ABC is overcoming the computational constraints posed by working with limited resource. The requirement to repeatedly simulate from a model can make inference extremely computationally expensive. 
Consequently, much of the methodological development in ABC has focused on improving computational efficiency, either through the use of more efficient Monte Carlo algorithms, or through the use of statistical methods to ameliorate the effect of  using a large tolerance.

The difficulty  of dealing with limited computer power is felt more keenly in climate science than in most other disciplines. 
A major focus of climate research concerns the construction of 
ever more accurate and comprehensive simulators of the climate system. Since the 1970's, global climate models have evolved from representing only the large-scale circulation of the global atmosphere 
\citep[e.g.,][]{Holloway1971simulation}  to models that incorporate complex dynamic representations of land surface, ocean, sea ice, atmospheric aerosols, ocean biogeochemistry, vegetation, soils and atmospheric chemistry \citep{IPCC2014models}.
Separate Earth system components are coupled through the exchange of fluxes, which describe the flow of some quantity between them (e.g. energy, moisture, $CO_2$), and by passing any state variables that are needed to define boundary conditions (e.g. land surface albedo, sea surface temperature).
 ``Intermediate complexity'' models (which use simplified model components and lower resolution in return for a more complete description of the Earth system and higher computational efficiency) may also include dynamic representations of other important elements, such as ice sheets, permafrost, ocean sediments and weathering \citep{IPCC2014models}, but these additional, long-timescale components require orders of magnitude longer simulations to reach equilibrium. 
Modern climate models are generally, and more accurately, described as ``Earth system models'' or ESMs.

This evolution in complexity has been accompanied by a 5-fold increase in spatial resolution, allowing the resolution of important finer scale processes. This increased resolution (combined with shorter time-steps that are required for numerical stability at higher spatial resolution) has \emph{alone} led to an $O(1000)$-fold increase in computational demands since the 1970's. In general higher resolution allows more direct and more realistic representation of smaller-scale processes, although this does not guarantee better projections, in part because more complex models are more challenging to calibrate. A feature of climate modelling is that multi-decadal climate projections must be used before data are available to validate them, while past data give only approximate clues to the expected behaviour of model discrepancy because expected changes greatly exceed the range of variability in the instrumental period.

It is perhaps inevitable, given the continual striving for more complex models and the highest possible resolution, that state-of-the-art ESMs will always be at the limits of what is practicable with available computing power. The UK Met Office Hadley Centre's computer comprises eight `supernodes' of IBM Power775 supercomputer servers, which were installed in 2012 at a cost of more than \pounds 11 million. The ESMs run at the Hadley Center and at equivalent climate modelling institutions in other countries are extremely computationally expensive, requiring months of such supercomputing to perform a single simulation of order 100 years.
Even the intermediate complexity model GENIE-1 \citep{holden2013model} used  in our case study (Section \ref{section:case}) requires several days (on a single CPU node) to perform each O(10 kyear) ``spin-up'' simulation to reach equilibrium, so that simulation ensembles require implementation on multi-node computing clusters.
The simulated climates are large complex datasets which comprise temporally-resolved three-dimensional spatial arrays of up to $\sim 100$ state variables. 
These outputs, in particular the outputs of carefully designed model inter-comparison projects, are often analysed in great detail, in a comparable way to how scientists in other fields analyse the outputs from empirical studies; model projections are the best, and only predictions we have  of future climate.

An ESM configuration is determined by the settings of many 100's of model parameters. These include switches (which determine the precise numerical schemes applied), physical constants that are approximately known but vary spatially in the real world (such as the reflectivity of ice) and parameterisations of ``sub-gridscale'' processes such as cloud formation, which have ``tuned'' values that are known to result in reasonable model behaviour.
This complexity (many weakly constrained inputs, high dimensional outputs and expensive simulators) has meant that careful statistical calibration (either with Bayesian or frequentist approaches) does not have a long history in climate science. Often different modules of an Earth system simulator are separately ``tuned'' before being bolted together. For example, the atmospheric component can be tuned independently of the ocean component by prescribing 
sea-surface temperatures with observational values. The components may all be tuned independently before being coupled, with no guarantee that what was a good tuning in an isolated module will work well in the coupled model. After coupling, a small subset of model parameters are adjusted so that the coupled model is consistent with large-scale observational constraints.
It has been shown, perhaps unsurprisingly, that such a tuning process does not produce a unique solution, so 
that different combinations of parameters can lead to equally plausible model realisations \citep{mauritsen2012tuning}. 

As statistical methodology develops, scientists are beginning to perform more careful parameter estimation in their models.  More rigorous parameter estimation methods are often developed with (relatively fast) intermediate complexity models, e.g., \citet{annan2005parameter}, thereby informing application to higher-complexity models, e.g., \citet{marquis2014investigation}.
We can  view climate simulators as black boxes which map from parameter values $\theta\in\Theta$, to climate states $f(\theta)=\MC_{\rm{sim}}$.
The aim of a Bayesian calibration, is to find the posterior distribution
\begin{equation}\label{eqn:calibpost}
\pi(\theta  | \MC_{\rm{obs}})\propto \int \pi(\MC_{\rm{obs}} | \MC_{\rm{sim}}) \pi(\MC_{\rm{sim}} | \theta) \rd \MC_{\rm{sim}} \pi(\theta),
\end{equation}
where $\MC_{\rm{obs}}$ is a set of observations of the climate system \citep{Kennedy_etal01, Rougier2007}. Here, $\pi(\theta)$ is the prior distribution for $\theta$,  $\pi(\MC_{\rm{sim}} | \theta) $ is the simulator likelihood function (which is typically unknown) and $\pi(\MC_{\rm{obs}} | \MC_{\rm{sim}})$ is the statistical model relating the simulator to physical climate. This calculation, however, is typically far too ambitious to perform in practice. 
Computational restrictions generally limit us to an ensemble of $N$ simulator runs $\{\theta^{(i)}, \MC_{\rm{sim}}^{(i)}\}_{i=1}^N$. Typically, $N$ is small, ruling out most Monte Carlo based calibration approaches. We are left needing to estimate $\pi(\theta | \MC_{\rm{obs}})$ as best we can, often by adding further approximation.

A further problem faced by climate scientists is that  
 simulator discrepancy (often called model error) can be considerable \citep{Murphy_etal2004}.  And whilst the physical models of climate, $\pi(\MC_{\rm{sim}}|\theta)$, are well developed,  statistical models of the simulator discrepancy relating simulated to observed climate, $\pi(\MC_{\rm{obs}}| \MC_{\rm{sim}})$, have only begun to be developed relatively recently \citep{Rougier_etal2014}.  The large simulator discrepancy makes most simulators incapable of  reproducing all aspects of  the climate record simultaneously and can mean that the simulator parameters are no longer directly comparable to their physical namesakes, making prior specification challenging.

So what is possible? We know that ABC, given infinite computational resource and a perfect simulator, can in theory produce arbitrarily accurate posteriors (i.e., the ABC posterior can be made arbitrarily close to the true posterior). But for many problems,  computational resource is often severely constrained and simulator discrepancy can be significant and largely unmodelled. Climate science is interesting for the statistician as it presents extreme cases of both these issues. 

A key idea allowing calibration in many of these expensive simulators is the idea of replacing the simulator with an emulator (or meta-model), which is a cheap statistical surrogate used in place of the simulator \citep{Sacks_etal89, Santner_etal2003, BACCO}.  Emulation techniques are attracting considerable interest in the climate community. They are used, for instance,   to approximate probabilistic model outputs \citep{sanso2008inferring,rougier2009analyzing,harris2013probabilistic}, for parameter estimation \citep{sham2012inferring,olson2012climate}, to facilitate model understanding \citep{lee2012mapping,holden2015emulation} and to provide numerically efficient model surrogates for coupling applications \citep{castruccio2014statistical,holden2014plasim,oyebamiji2015emulating}.
The application we will describe here is in the ABC design of ``plausible'' simulation ensembles \citep{holden2010probabilistic,edwards2011precalibrating}, using emulation in order to overcome the prohibitive limitations imposed by simulator cost.
 
 \section{History-matching and ABC}
 
Climate science presents the double whammy of computationally expensive simulators, and simulator discrepancy that is too large to ignore but which is not well understood or modelled. Both of these issues make a careful Bayesian calibration (as described by Equation \ref{eqn:calibpost}) difficult. What can be achieved? Our aim is to compare observations of Earth's climate $\MC_{\rm{obs}}$, with simulator predictions $\MC_{\rm{sim}} = f(\theta)$, in order to learn about the parameter $\theta$. ABC is an approach for obtaining a probabilistic calibration, and seeks to match simulator output to observations, approximating the distribution
\begin{equation}\label{eqn:ABC}
 \pi_{ABC}(\theta | \MC_{\rm{obs}}) \propto \int \mathbb{I}(\rho(\MC_{\rm{obs}}, \MC_{\rm{sim}})\leq \epsilon) \pi(\MC_{\rm{sim}}|\theta) \rd \MC_{\rm{sim}} \pi(\theta).
 \end{equation}
 The acceptance kernel $\mathbb{I}(\rho(\MC_{\rm{sim}}, \MC_{\rm{obs}})\leq \epsilon)$ implicitly implies a uniform distribution for the simulator discrepancy \citep{Wilkinson_2013},  but this is usually viewed as a pragmatic compromise, rather than a modelling decision.

An alternative to a probabilistic calibration, is to do a history match \citep{williamson2013history}, which has been used in studies involving complex computer models, such as oil reservoir
modelling \citep{Craig_etal1997},  cosmology \citep{Vernon_etal2010}, epidemiology \citep{Andrianakis_etal2015}, and climate science \citep{edwards2011precalibrating}.
 History matching, like calibration, seeks to identify regions of the input space that give acceptable matches between simulator output, $\MC_{{\rm sim}}$, and observed data, $\MC_{\rm{obs}}$. 
But instead of finding a probability distribution over $\Theta$,  we instead seek merely to rule out implausible regions of input space, i.e., those $\theta$ that the simulator suggests could not have lead to $\MC_{\rm{obs}}$, even after having accounted for the simulator discrepancy.
Often large parts of the input space give simulated climates that are very different from the observed data, and which can hence be ruled to be physically implausible and removed from further consideration. 

We define $\MP_\MC$ to be a set of plausible climate states that represent an acceptable match between simulation and observation. We define $\MP_\theta$ to be the subset of the parameter space that leads to plausible simulated climates, i.e., 
$$\MP_\theta = \{\theta \in \Theta : f(\theta) \in \MP_\MC\}.$$
Often,  the vast majority of the input space  gives rise to unacceptable matches to the observed data (sometimes $\MP_\theta=\varnothing$), and it is these regions that we are trying to rule out as implausible.
 For example, for an ESM, we might define $\MP_\MC$ to be any simulated climate that has global surface air temperature (SAT) within $2\degree$C of the observed value, the 
maximum value of 
Atlantic Meridional Overturning Circulation (AMOC), a measure of the large-scale circulation of the ocean, within 5 Sv (1 Sv $= 10^6 m^3 s^{-1}$) of observations, and the global mass of vegetation to be within 200 giga-tonnes carbon (GTC) of observations, though clearly the choice of appropriate metrics and acceptance ranges is highly simulator-dependent. $\MP_\theta$ is then the set of model parameters that would generate plausible climates for the ESM in question.
 
 Note the similarity to ABC here. If the prior distribution for $\theta$ is uniform on $\Theta$, i.e.,
 $\pi(\theta) \propto \mathbb{I}_{\theta \in \Theta}
 $,  if $f(\theta)$ is deterministic (as is often, at least approximately, the case in climate science), and if we use $\mathbb{I}_{f(\theta) \in \MP_\MC}$ as the ABC acceptance kernel, then
 $$\pi_{ABC}(\theta | \MC) \propto \begin{cases}1 \mbox{ if }\theta \in \MP_\theta\\
 0 \mbox{ otherwise}.
 \end{cases}
 $$
If we interpret a posterior probability of zero, as the statement that $\theta$ is implausible, then  history matching and ABC are thus the same.
 Note also the direct relationship between the discrepancy considerations  built into $\MP_\MC$, and the way ABC  performs `Monte Carlo' exact inference for the model that has a discrepancy defined by the acceptance kernel \citep{Wilkinson_2013}.
 
History matching and ABC have in common that they do not use a detailed model of the discrepancy, but instead characterise it using simple criteria.
 A philosophical difference between the two approaches perhaps lies in the degree of thought given to the plausible set $\MP_\MC$. In history matching, the plausibility 
 criteria are often based on measurement error variances and the expected magnitude of the simulator discrepancy \citep{Vernon_etal2010}. Consequently,  $\MP_\theta$ 
consists of those 
parameter values $\theta$ that have not yet been ruled out as implausible by our knowledge of the simulator and its discrepancy, and the observed data and  measurement error. The result is usually not interpreted  probabilistically, but only as values that we can not yet rule to be implausible given our current state of knowledge.
In contrast,  in ABC the choice of metric $\rho$ and tolerance $\epsilon$ 
are usually based pragmatically on the characteristics of the algorithm, rather than on physical aspects of the underlying problem.
Often, $\epsilon$ is chosen to generate a specified number of acceptances. For example, if the computational budget allows for $10^8$ simulator runs, and we want $10^4$ accepted values in order to approximate the posterior, we set $\epsilon$ to the value that leads to 0.01\% of simulations being accepted \citep[for example,][interpret ABC as a nearest neighbour algorithm]{Biau_etal2015}. 

 A further difference lies in the choice of information to include in $\MP_\MC$ (i.e., what summary statistics to use). Climate simulators provide a large variety of outputs, and some of these are 
better able to reproduce observed climate than others. For example, temperatures are generally better reproduced than precipitation, consequently, it is more common to calibrate to the former than the latter.  In contrast, ABC 
has its roots in genetics, where perhaps the simulator output is less varied, and consequently, more focus is given to the automatic selection of summary statistics,  often chosen on the basis of what is most informative for $\theta$ \citep{Blum_etal2013}.  This approach is unlikely to be suitable in climate science. Some outputs for which the simulator discrepancy is particularly large (precipitation say) may well be very `informative' about $\theta$ if we do not allow for discrepancy, but this would only  misguide and may lead   us to incorrectly rule out large swathes of parameter space as implausible. 
Variables which are not well simulated are often included in ESMs, either because they improve the overall simulation through the representation of important feedbacks, or because they are considered important outputs in their own right in spite of higher discrepancy associated with the outputs. Whether a weak calibration constraint on these ouputs is appropriate will depend on the details of the discrepancy. Where a known missing process gives a significant contribution to regional error for instance, such as large precipitation errors in monsoon regions as a result of unresolved topographic variation, using a too precise calibration constraint (equivalently too small a model discrepancy)  would distort the rest of the solution. 
  
 %
%
%
%

 A key question for any simulator is whether given a  set of plausibility conditions, the simulator is capable of producing any plausible simulated climates. That is, is $\MP_\theta$ empty? 
 If $\MP_\theta$ is empty, it is an indication that we understand less than we thought about the simulator and system. Either there is an error in our implementation of the simulator, or we have under-estimated the magnitude of the simulator discrepancy or measurement error. 
 The fact that the result of a history-match can be to find there are no plausible parameter values  should not be seen as a negative aspect of the approach, as  it forces us to confront the cold reality that something is missing from our understanding of the system. In contrast, 
 likelihood based 
techniques such as MCMC (and  pragmatic ABC applications where $\epsilon$ is chosen to guarantee a particular acceptance rate), result in a posterior distribution $\pi(\theta | \MC_{\rm{obs}})$ regardless of how close the simulated climates are to real climate. It is thus sensible when using these techniques, to carefully check that the calibrated simulator does indeed produce acceptable fits.
While it can often be useful to find the distribution $\pi(\theta | \MC_{\rm{obs}})$ (or an approximation to it) regardless of the simulator quality, note that  
  if discrepancy is ignored, $\pi(\theta | \MC_{\rm{sim}})$  can often be more constrained, or equivalently $|\MP_\theta|$ smaller, than is justified  \citep{Brynjarsdottir_etal2014}. 
  
Note that even if a probabilistic calibration is required, a history match  can be performed first in order to rule out regions of space which are clearly implausible. This can  dramatically reduce the area needed to be explored during  the more challenging probabilistic calibration. If using a stochastic simulator, for which $\theta$ may never be completely ruled as implausible (as $\pi(\theta | \MC_{\rm{sim}})>0$ for all $\theta$ say), this can still be advantageous. We can rule out parameter regions for which the likelihood is considerably smaller than at the MLE with only a small increase in the approximation error \citep{Wilkinson2014}, again 
making a subsequent probabilistic calibration easier. 
 
 Assuming that $\MP_\theta$ is not empty, the question then becomes, can we find elements of $\MP_\theta$,  and better still, can we characterise all of $\MP_\theta$? 
The complexity of climate science is such that even incomplete specifications of $\MP_\theta$ are useful, as discussed in Section \ref{section:case}. This is because interest lies not in $\MP_\theta$, but in what it implies about future climate, i.e., in the implied calibrated distribution for other aspects of the climate system 
$$\pi(\MC_{\rm{future}}  | \MC_{\rm{obs}}) = \int \pi(\MC_{\rm{future}} |\theta) \pi(\theta | \MC_{\rm{obs}})d\theta.$$
and so even partial descriptions of $\MP_\theta$ are useful in constraining our beliefs about future climate behaviour.
Our aim is thus, given a limited computational budget of $N$ simulator evaluations, can we find $\MP_\theta$ and the corresponding set of  plausible future climates? ABC applications usually use millions of simulator evaluations. What can we do if instead we can only afford 100 or 1000 simulator evaluations? 
 The answer is going to be even more approximate than in ABC, and furthermore, we will necessarily  have to make some modelling assumptions if we wish to make progress. The key tool that has arisen for doing this is the 
 emulator, or meta-model.

 \section{Emulation}\label{sect:emulation}
 
If the simulator, $f(\theta)$, is expensive to evaluate, we can instead try to find an approximation, $\tilde{f}(\theta)$, called an emulator or meta-model, which provides a good approximation to $f(\theta)$, but which is computationally cheap \citep{Sacks_etal89, BACCO}.
We can then either use $\tilde{f}$ to answer the question of interest (e.g., calibrating the simulator), or use it to guide the choice of the next parameter value at which to evaluate $f$.

We  start by generating an ensemble of simulator evaluations $\CD=\{\theta_i, f(\theta_i)\}_{i=1}^N$, which we then use to build $\tilde{f}$. 
Building an emulator is a regression problem, and consequently a myriad of different techniques have been used, including linear regression and its variants,  neural networks, and Gaussian processes (GPs, also known as Kriging), with  GPs proving the most popular class of model thus far. The functional form of the simulator is not known a priori, and so neither is the best regression model, but  a reasonable approximation can usually be found using GPs, as long as the response is a smooth continuous function of $\theta$. For the purposes of calibration, the key properties of any emulator are predictive accuracy, quantification of uncertainty in the predictions, and  speed of prediction. In climate science, where the output fields being modelled are often spatio-temporal fields, the regression model is usually combined with a dimension reduction technique, to project the output onto a lower dimensional manifold \citep{Higdon_etal2008, holden2010dimensionally, Wilkinson2010}.

 \subsection{Sequential history matching}

For many problems, the plausible set $\MP_\theta$ may constitute only a small fraction of the prior space $\Theta$. Furthermore, $\MP_\theta$ may consist of multiple disconnected regions. For Monte Carlo methods, this can make designing an effective sampler difficult, as MCMC chains (or particles) can fail to explore all plausible regions. For emulator methods, the difficulty lies in building a model that can approximate the simulator in all regions of space. For example, stationary covariance functions that assume a constant length-scale throughout space are commonly used in GPs, and may  be inappropriate. Other problems arise if $f(\theta)$ varies over too wide a range, which is common if $f(\theta)$ is a likelihood function \citep{Wilkinson2014}.
If we need to an  emulator of $f(\theta)$ that is valid in all of $\Theta$, then we can look to use a non-stationary covariance function or a more flexible model such as a treed-GP \citep{Gramacy_etal2008}.
 However, for calibration, we only need approximate the simulator when $f(\theta)$ is close to being plausible. In other parts of parameter space, it is only necessary to say $\theta$ is implausible with a high degree of confidence. It does not matter if an estimate of $f(\theta)$ is poor, as long as we are correct in saying $f(\theta) \not \in \MP_\MC$.

GP predictions are more accurate in regions rich in data. Thus, the key issue when building a GP emulator is the choice of the design points, $\CD_\theta=\{\theta_i\}_{i=1}^n$, at which we evaluate the simulator.  Space filling designs, such as maximin Latin hypercubes \citep{Mckay_etal2000} or low discrepancy sequences \citep[such as Sobol sequences,][]{Morokoff_etal1994} are the default  choice of design, and usually lead to reasonable global approximations. But they are less well suited to calibration problems, 
in which we usually want to focus on just a small region of parameter space.

Instead of a space filling design, we can seek to build  the design sequentially: given the current design, we build an emulator that describes our current knowledge of $f(\theta)$. We then use the emulator to decide where next to run the simulator, and so on.
 The basic idea is as follows:
 \begin{enumerate}
 \item Start with an {\it a priori} plausible set $\MP_\theta^{(0)} = \Theta$.
 \item Choose design $\CD_\theta^{(1)}= \{\theta_i \in \Theta: i=1, \ldots, n_1\}$, and  run the simulator, to get ensemble $\CD^{(1)} = \{(\theta_i, C_i = f(\theta_i): \theta_i \in \CD_\theta^{(1)}\}$.

\item Build emulator $\tilde{f}_{(1)}$ and use it to predict the plausible set $\tilde{\MP}_{\theta}^{(1)}$.

\item Choose new design points $\CD_\theta^{(2)}$, and run the simulator to get $\CD^{(2)}$.
\item Build emulator $\tilde{f}^{(2)}$ and use it to predict the plausible set $\tilde{\MP}_{\theta}^{(2)}$.
\item Etc.
 \end{enumerate}

The details of each step vary in each problem. 
The plausibility 
criteria are usually defined so that they become more stringent at each iteration. The first plausibility condition $\MP_\MC^{(1)}$ may be relatively weak, with $\MP_\MC^{(1)}, \MP_\MC^{(2)}, \ldots, \MP_\MC^{(W)}$ slowly approaching the final desired criterion $\MP_\MC^{(W)}$.  If the difference between $\MP_\theta^{(i)}$ and $\MP_\theta^{(i+1)}$ is too large, we may find the emulator accuracy is insufficient, causing us to incorrectly  rule-out some regions of space (type-I errors). The plausibility criteria can be relaxed  by changing the number of measurements we need to match, and the closeness of the required match. Note 
 the superficial similarity to SMC-ABC approaches, in that the approximation  is iteratively improved as we learn.

The emulator used at each stage may be based upon all the previous simulator runs, adding new data points in important regions (see below for details), or it can be built from scratch. For example, in \citet{Vernon_etal2010} they build an emulator, $\tilde{f}^{(i)}$, to predict $f(\theta)$ for $\theta \in \MP_\theta^{(i-1)}$, the estimated plausible region from the previous iteration. The emulator is not required to predict for $\theta \not \in \MP_\theta^{(i-1)}$. The benefit of this is that the simulator response is likely to be less variable within $\MP_\theta^{(i-1)}$ than in $\Theta$, making it easier to model. The disadvantage is that  if some regions are incorrectly ruled to be implausible in iteration $i-1$, this mistake can never be rectified.

The most important algorithmic decision is  the choice of design, $\CD_\theta^{(i)}$, at each iteration, i.e., given an emulator, how should we choose locations $\theta$ at which to run the simulator? If we 
only wish the emulator to predict well in $\MP_\theta^{(i-1)}$, then we only need a design in $\MP_\theta^{(i-1)}$. 
\citet{Vernon_etal2010} take the approach of seeking to use a space filling design on $\MP_\theta^{(i-1)}$, such as a Latin hypercube. To do this, they create a large design on $\Theta$ and then reject any point not predicted to lie in $\MP_\theta^{(i-1)}$ by $\tilde{f}^{(i)}$, which is also the approach we describe in Section \ref{section:case}. 
If we instead seek a global emulator valid for all $\theta \in \Theta$, but which  is accurate in the important regions, then it can be beneficial to  add simulator runs to the design one at a time. 
The critical regions are those where the emulator is most uncertain about whether $\theta \in \MP_\theta$. This is typically either in regions for which we have no data, or near the edge of the plausible region where we are unsure if $\theta \in \MP_\theta$ or not given the accuracy of the emulator.
If we use a GP emulator, then our prediction of $f(\theta)$ is Gaussian:
$$f(\theta) \sim N(\mu^{(i)}(\theta), \Sigma^{(i)}(\theta))$$
where $\mu^{(i)}$ and $\Sigma^{(i)}$ are the mean and covariance function of $\tilde{f}^{(i)}$. This allows us to calculate the probability that $\theta \in \MP_\theta$. For example, if our  criterion is that $\theta $ is plausible if $D_- \leq f(\theta) \leq D_+$, then 
$$p(\theta) = \BP_{\tilde{f}^{(i)}}(\theta \in \MP_\theta) = \Phi\left(\frac{D_+ - \mu(\theta)}{\Sigma(\theta)^{\frac{1}{2}}}\right)- \Phi\left(\frac{D_- - \mu(\theta)}{\Sigma(\theta)^{\frac{1}{2}}}\right).$$  
In some regions $p(\theta)$ will be close to zero, indicating that we are confident that $\theta$ is implausible, and in others close to one, indicating the converse. It is  regions in  which we are most uncertain, that we wish to target, as these 
represent parameter values that we can neither rule in nor out. One approach to selecting new design points is to choose points to minimize the entropy of this surface \citep{Hennig_etal2012, Chevalier_etal2014}. The entropy represents how close to certain knowledge we are. If we let $\bar{H}$ be the average entropy of the emulator prediction of the plausibility surface:
$$\bar{H} = \int -p(\theta) \log p(\theta) -(1-p(\theta)) \log (1-p(\theta)) {\rm d} \theta$$
then we can ask, if we were to add a simulator evaluation at $\theta$, what is the expected value of $\bar{H}$ given the expected 
resulting information? We can then add $\theta$ to the design in order to minimize $\BE( \bar{H} | \CD^{(i-1)}\cup \{\theta\})$. This approach places new points in regions that 
most quickly  
reduce the uncertainty about the plausible region $\MP_\theta$. 
 
 \subsection{A simple climate example}\label{section:toy}
 
As an illustration of the potential benefit of these techniques, we consider a relatively simple two-box climate simulator \citep{emanuel2002simple}, which models atmospheric and ocean heat transport and storage, with water vapour as a positive feedback. The simulator is useful for the purposes of demonstration, as 10 years of model time takes approximately 5 seconds of CPU time, allowing a large number of model runs to be done.
Matlab code for this simulator is available online.

We present the results of a simple history-matching task, calibrating  two parameters: DTcrit\_conv, the critical vertical temperature gradient that triggers convection, which we allowed to vary in
the range $[30,50]$; and GAMMA, the emissivity parameter for water vapour, which we varied in 
the range $[1,2]$. We try to find the parameter values that give a global surface temperature  between 294.5K and 295.5K once the model is in equilibrium. The CO$_2$ concentration was set to 560ppm, and all other parameters were set to their default values \citep{EPmanual}. These choices are arbitrary and only intended for illustration of the methodology. 

\begin{figure}
 \begin{center}
\includegraphics[scale=0.4, bb=120 0 1500 500, clip]{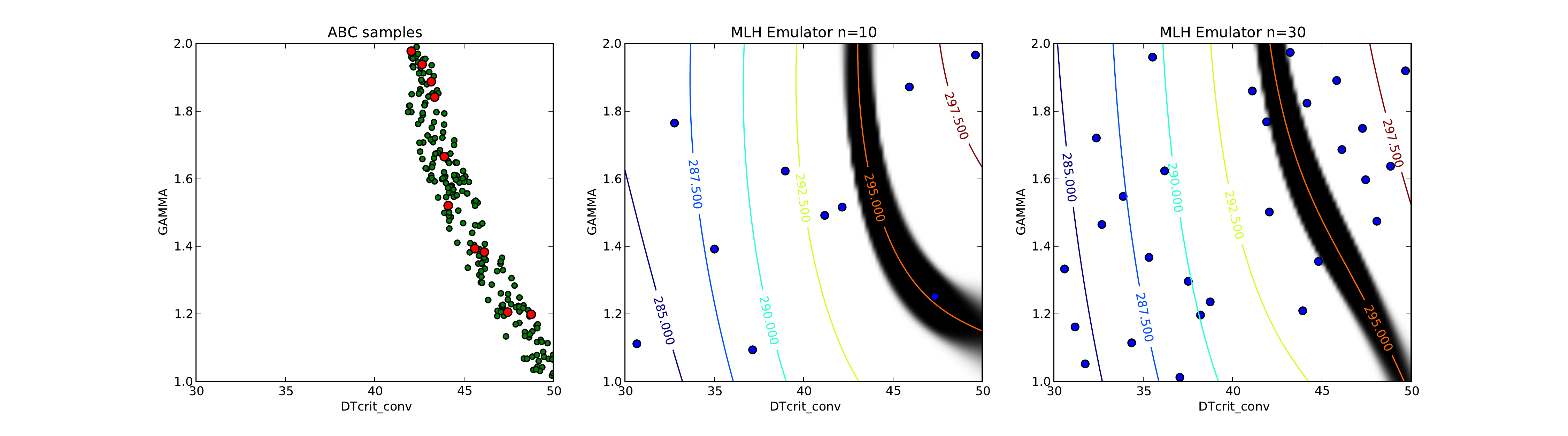}
 \end{center}
\caption{Left: Accepted samples from the rejection ABC algorithm after 100 (red) and 1000 (green) simulator evaluations. Middle and right: The estimated plausible region using an emulator trained with a maximin Latin hypercube design (points shown in blue) with 10 (middle) and 30 (right) simulator evaluations. The shading indicates the estimated value of $\BP(\theta \in \MP_\theta)$. The contour lines are the estimated response surface $f(\theta)$.  }
\label{fig:ABC}
\end{figure}

Applying a simple ABC rejection algorithm  and allowing for 1000 simulator evaluations gave us 106 accepted parameter values, which are shown in the left-hand plot in Figure \ref{fig:ABC}, with the red points showing the 10 values accepted after only 100 simulator evaluations. In contrast, the middle and right-hand plots 
shows the result of using a GP emulator with a maximin Latin hyper-cube (MLH)  design of 10 and 30 simulator evaluations. After 10 simulator evaluations, the emulator has some idea of where the plausible region is, but with errors for example in the bottom right hand corner. After 30 simulator evaluations it has  accurately found all of the plausible region, with just a little uncertainty at the edge of the region (shown by the grey shading). 

This approach, however, relies upon finding a good design. If  an accurate emulator results, then it will do well at predicting the plausible region. 
Here we can see, in the case where we had only 10 design points, that no information is available about the bottom right hand corner of the parameter space, and consequently the model does less well 
there. As the design is chosen in advance of the simulations being run, finding a good design involves an element of luck.

\begin{figure}
%
\includegraphics[scale=0.4]{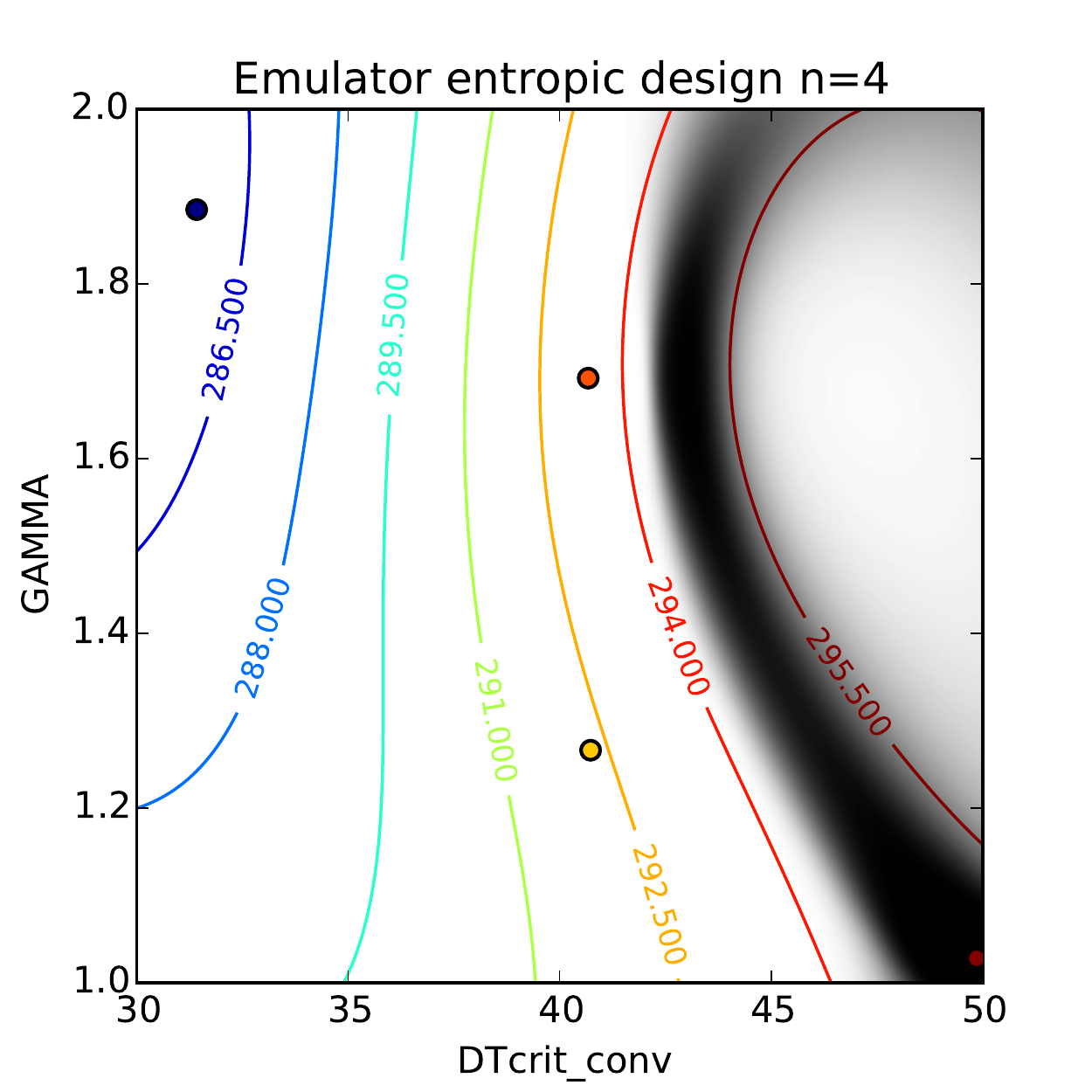}\includegraphics[scale=0.4]{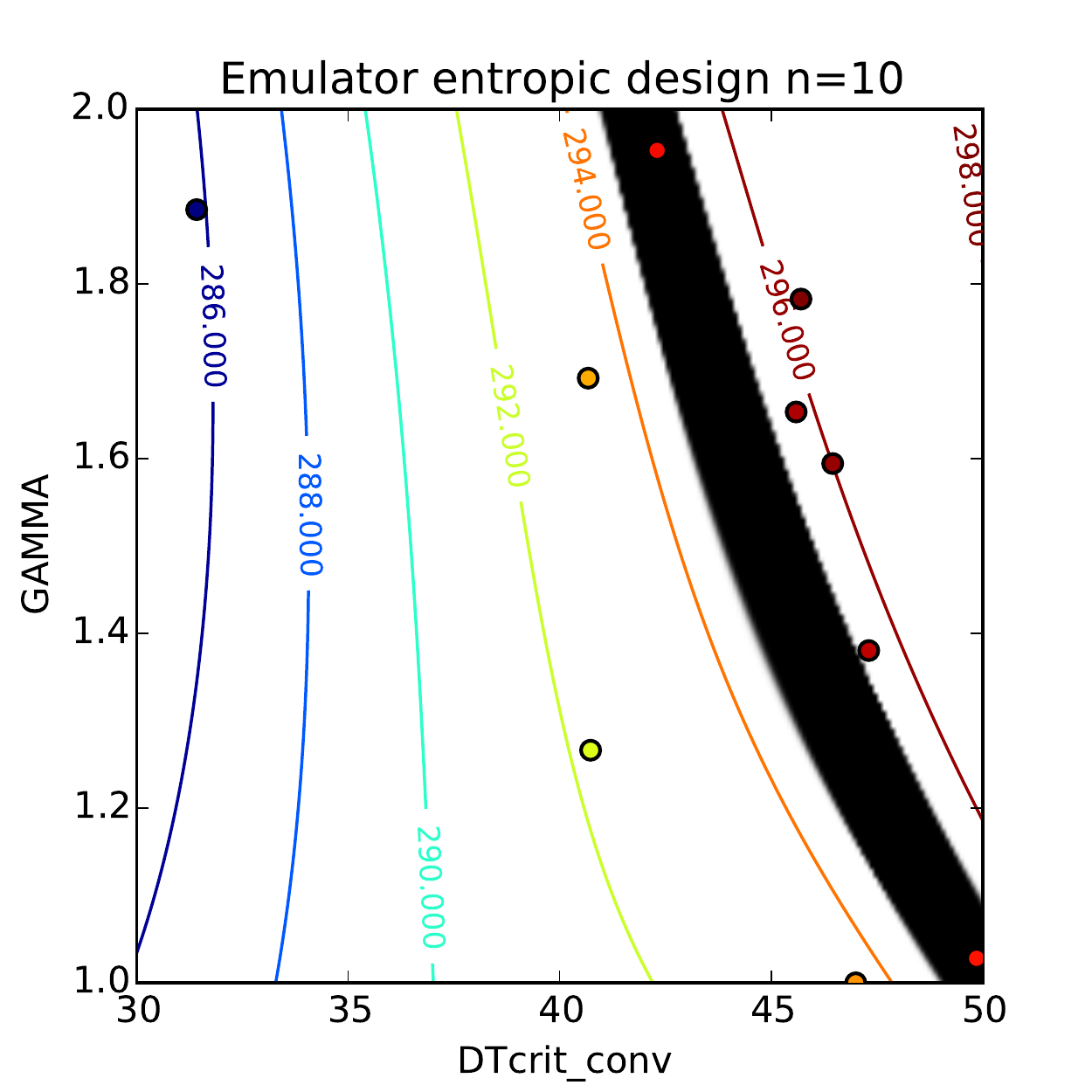}
\caption{Results from using an entropy based sequential design. The left-hand column shows the estimated response surface (contours) and $\BP(\theta \in \MP_\theta)$ (shading), with the design points overlaid. The large red point is the most recently added point. The right-hand column shows the entropy surface. The top row uses four simulator evaluations, and the bottom row uses 10 simulator evaluations, all added according to the entropy criterion.}
\label{fig:emulation}
\end{figure}

If we instead use a sequential design and add  design points one at a time in order  to minimize the expected average entropy of the resulting history-match, then we can significantly improve the speed with which we find $\MP_\theta$. 
The two plots in Figure \ref{fig:emulation} show the resulting history match after 4 and 10 simulator evaluations. After only 10 simulator evaluations, we have found $\MP_\theta$ with superior accuracy to that found after 30 simulator evaluations using the MLH design. 

Note that the acceptance rate in the ABC algorithm was approximately 10\%, considerably higher than in most problems (we had  a 1\%  acceptance rate in the case study described in Section \ref{section:case}). As the acceptance rate decreases, the value of using an emulator to predict $\MP_\theta$ increases, as the emulator is able to predict where the plausible region is, whereas ABC can only find the region by chance, as it uses no information about the shape of the underlying surface. 
In contrast, the major advantage of the Monte Carlo approach is that it is less prone to errors (although mixing errors commonly occur in practice), unlike the emulator approach, which can mislead if the fitted model is inaccurate, and thus requires careful supervision.


 
 \section{Climate model case study}\label{section:case}
 
\subsection{The global carbon cycle}

Human emissions of 
CO$_2$ into the atmosphere 
are a principal cause of climate change. However, this ``anthropogenic'' CO$_2$ does not remain in the atmosphere indefinitely. It is taken up by vegetation and by the oceans, and eventually (after many thousands of years) it is deposited as carbonate sediments at the ocean floor. Understanding these processes is crucial for future climate projections. Climate change is driven by changes in CO$_2$ concentration and it is therefore determined by the interplay between anthropogenic emissions and the carbon cycle. Many carbon cycle processes are highly uncertain. Projections of 
year 2100 CO$_2$ concentrations from different Earth System Models (ESMs) driven by the same assumption of future emissions typically vary by $\pm100$ ppm \citep{friedlingstein2006climate}. This uncertainty range is greater than the total increase 
to date (2015) due to all historical anthropogenic emissions ($\sim120$ ppm) 
 
To investigate uncertainties in the global carbon cycle we need a model of appropriate complexity
that is capable of resolving the important processes but which is sufficiently numerically efficient. The GENIE-1 intermediate complexity ESM \citep{holden2013model} is one such model. The computational speed of GENIE-1 comes mainly from the use of a very simple 2D model of the atmosphere and relatively coarse model resolution (grid cells of $\sim$~1,000 x 1,000 km). The carbon cycle of GENIE-1 comprises a terrestrial carbon model, a 3D dynamic ocean, dynamic sea ice, ocean biogeochemistry and ocean sediments. Given appropriate model parameter choices, GENIE-1 simulates realistic spatial distributions of carbon storage in vegetation, soil, ocean and carbonate sediment. However, the future response of the climate cycle to ongoing emissions depends upon the specific parameter choices, and will vary even amongst parameter sets that have been constrained to produce similar (and reasonable) modern climate states. To quantify this uncertain response we require an ensemble of simulations that samples widely from plausible input parameter space.

The timescales for different carbon cycle processes vary considerably. Equilibrium timescales are $\sim10$'s years for vegetation, $\sim100$'s years for soil, $\sim1,000$'s years for the ocean and $\sim10,000$'s years for carbonate sediments. In order to simulate an Earth with a carbon cycle in approximate equilibrium (i.e. prior to human interference), a simulation of at least 10,000 years is required\footnote {Shorter spin-ups are sufficient for models that neglect sediments.}.  Although 
several orders of magnitude faster than than state-of-the-art ESMs, GENIE-1 requires $\sim$ 4 CPU days to simulate 10,000 real years. 
The exploration of high-dimensional input space and identification of plausible 
subspaces is therefore a highly demanding computational problem, which we address through emulator-informed ABC.
 
\subsection{Emulator-informed ABC design}

The philosophy of the design approach is to vary key model parameters over the entire range of plausible values and to accept those parameter combinations that lead to climate states that cannot be uncontroversially ruled out as implausible \citep{edwards2011precalibrating}. We are seeking to explore all plausible simulator realisations  in order to capture the range of possible feedback strengths. The input ranges we apply, $\Theta$, are generally broader than ranges that are applied in model tuning exercises. This is in part to enable us to fully quantify model behaviour over plausible parameter space, $\MP_\theta$, and in part to improve the validity of the ensemble for application to diverse climate states, such as the Last Glacial Maximum.

The experimental set-up is described in \cite{holden2013controls}. We varied 24 model parameters in the ensemble. The choice of parameters was governed by consideration of the processes that are thought to contribute to the natural variability of atmospheric CO$_2$ on glacial-interglacial timescales \citep{kohfeld2009glacial} and hence to which the distribution of carbon may be sensitive in general.  Five atmospheric parameters were varied. These parameters control the spatial distribution of simulated temperature and precipitation, and hence drive changes in vegetation, sea-ice coverage and ocean circulation. Five parameters were varied in the vegetation model, controlling photosynthesis and respiration rates. Five ocean parameters were varied. These control ocean circulation, and hence the spatial distribution of carbon, alkalinity, dissolved oxygen and nutrients in the ocean. Sea-ice diffusivity was varied, primarily because of its effect on ocean circulation by altering the transport of freshwater. Nine ocean biogeochemistry parameters were varied. These parameters drive changes in the rates of atmosphere-ocean gas exchange, plankton photosynthesis and the remineralisation of the organic products of this photosynthesis. The rate of remineralisation controls the transport of carbon from the surface of the ocean to the deep. 

A 500-member ensemble of 25,000-year simulations was first performed using a maximin Latin hypercube (MLH) design\footnote{We note that while efficiencies can be gained in certain applications by initialising each ensemble member with output from an existing equilibrium simulation, such an approach is not likely to be useful here as our approach is designed to sample widely differing Earth system states.}. The plausibility of each  simulator run was evaluated using eight different output quantities, usually termed {\it metrics} in the climate literature. These simple metrics impose no constraints on the spatial distribution of modelled outputs. They instead provide global-scale constraints on atmosphere (global average temperature), ocean (strength of North Atlantic overturning and Antarctic Deep Water formation), Antarctic sea-ice coverage, global vegetation carbon, global soil carbon, ocean biogeochemistry (average dissolved oxygen concentration in the global ocean) and ocean sediments (the average percentage of CaCO$_3$ in the surface sediment).  Only four of the 500 MLH simulations were found to satisfy all eight plausibility constraints, which given that $\dim \theta = 24$, is insufficient for any meaningful statistical analysis. The MLH ensemble took more than ten years of computing to complete, demonstrating that a naive application of ABC is infeasible for this application.

As described in Section \ref{sect:emulation}, we can  use emulators to guide the search to find plausible  regions of parameter space. Regression-based emulators, including linear and quadratic terms, were built for each of the eight metrics (outputs) specified above. Prior to fitting, variables were linearly mapped onto the range [-1,1] so that odd and even terms were orthogonal, aiding variable selection. The models were built using a stepwise model selection scheme, initially using the  Akaike Information Criterion as the selection criterion, and then subsequently shrunk further by applying the more stringent Bayes Information Criterion. This procedure of first growing the model beyond the BIC constraint and then shrinking 
helps to avoid local minima in the stepwise search.

Parameters were then sampled uniformly from the a prior plausible region and the emulators used to predict if they would lead to plausible simulations. Parameters were accepted as potentially plausible when the emulators predicted plausible values for all eight metrics. The plausibility ranges used were based on the observed climate record, the simulator discrepancy, and the emulator accuracy. Each accepted parameter combination was then used as a design point in a further simulation.

As simulations completed, the emulators were rebuilt using the additionally available data. This process progressively improved the success rate of the emulator predictions  (i.e. the percentage of emulator predicted plausible parameters that led to plausible simulations) from 24\% to 65\%. In total, the simulator was run for 1,000 parameter values predicted to be plausible by the  emulator. This produced 885 completed simulations of which 471 were plausible (the remaining 115 simulations terminated before completion, a common occurrence with climate simulators). This 471-member plausible set forms the Emulator Filtered Plausibility Constrained (EFPC) ensemble. The generation of these simulations required a further $25$ years of computing time. Without the $\sim 50$-fold increase in efficiency gained by using an emulator to predict the plausible region, this would have required an infeasible amount (more than 1000 years) of CPU time.

\begin{figure}
\begin{center}
\includegraphics[scale=0.6]{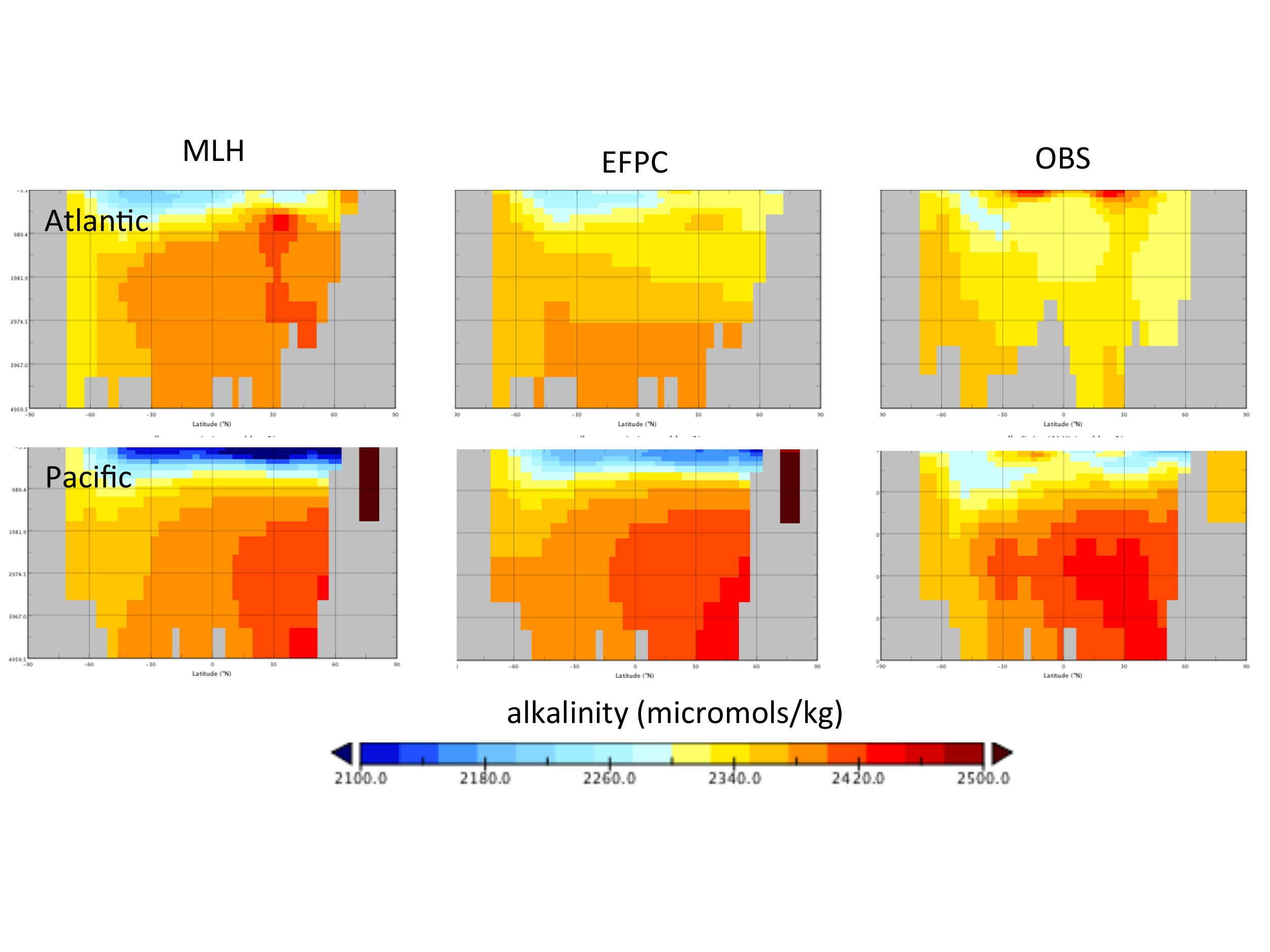}
\end{center}
\caption{Cross-sections of ocean alkalinity through the Atlantic ($25\degree$W) and Pacific ($155\degree$W) oceans. The figure compares the mean of the training MLH ensemble (left panels) and the plausibility filtered EFPC ensemble (centre panels) with observations (right panels).}
\label{fig:ALK}
\end{figure}

While it is clear that ABC strongly constrains the  outputs (metrics) that are explicitly filtered for, it is worth noting that it indirectly constrains all aspects of the Earth system and leads to improved simulated magnitudes {\em and spatial distributions} of state variables generally. Figure \ref{fig:ALK} provides an illustrative output of the EFPC ensemble, and of the benefits of the ABC filtering. The figure illustrates cross-sections of ocean alkalinity through the Atlantic and Pacific oceans, comparing ensemble means of the unfiltered MLH simulations (left) and filtered EFPC simulations (centre) with observational data (right). Ocean alkalinity exerts a strong control on atmospheric carbon dioxide by determining the degree to which dissolved carbon dioxide is dissociated into bicarbonate and carbonate ions, in turn determining the rates of exchange of dissolved carbon in the ocean with the atmosphere (carbon dioxide) and the sediments (calcium carbonate). Alkalinity is not directly constrained by the ABC metrics, but its distribution is influenced by them, for instance through the constraints imposed on ocean circulation strength and the sediment carbonate concentration. Relative to the MLH ensemble, the EFPC ensemble shows elevated surface concentrations, decreased concentrations in the deep Atlantic (apparently associated with the Atlantic overturning circulation in the unfiltered ensemble) and increased penetration of high alkalinity towards Southern latitudes in the deep Pacific. Although discrepancies with observations remain, which may reflect structural deficiencies in the simulator, each of these trends produces better ensemble-averaged agreement with the observed distribution.

\subsection{Applications} 

Although the use of ABC to derive a posterior distribution is useful in itself, 
our primary motivation is to identify a set of plausible  parameters for application to diverse simulation problems. The EFPC parameter set has been used in a range of experiments, considering both past and future climate change.  For clarity, it is worth emphasising that while these experiments did not use ABC directly, they were all rendered tractable by the use of emulator-informed ABC to design the underlying simulation ensemble. A selection of these experiments are summarised below, each with a focus on a different category of application.

\subsubsection{Probabilistic simulation outputs: the uncertain response of the carbon cycle to anthropogenic CO$_2$ emissions}

We (PBH and NRE) contributed a suite of carbon cycle experiments for the Fifth Assessment Report (AR5) of the Intergovernmental Report on Climate Change (IPCC). Fifteen Intermediate Complexity ESMs from around the world performed these experiments. The focus was on historical change \citep{eby2013historical} and long-term future change \citep{zickfeld2013long}, considering long timescale problems that are not tractable by state-of-the-art ESMs, thus requiring the use of reduced complexity models such as GENIE-1. Forty seven experiments were performed.

We applied a subset of the EFPC parameter set, in part to aid computational tractability,
in view of the 47 separate experiments required, 
and in part to eliminate a bias in the transient response of the ensemble. The EFPC parameter set is constrained to simulate a plausible preindustrial climate, but no constraint was imposed upon the dynamic response to anthropogenic emissions. Four important model parameters were not constrained by preindustrial plausibility, two relating to cultivated vegetation (deforestation for agriculture was neglected in the preindustrial simulations), a parameter controlling the direct effect of CO$_2$ on photosynthesis (``CO$_2$ fertilisation'', see following section) and a parameter controlling the uncertain effect of  clouds on the Earth's radiation budget in a warmer planet. The dynamic response was therefore filtered through a historical forcing experiment, which imposed anthropogenic forcing, including CO$_2$ emissions, since preindustrial times in an EFPC ensemble of transient simulations. Twenty parameter sets, selected at random from the EFPC parameter sets, but constrained to approximately reproduce the present day atmospheric CO$_2$ concentration, were accepted and applied to the IPCC experiments.

We do not attempt to summarise the results of these extensive multi-model comparisons here, but note that the GENIE-1 perturbed-parameter ensemble was found to provide an unbiased representation of the multi-model ensemble, being approximately centred on the mean of the fifteen models and with comparable uncertainty. These uncertainties were presented in a related model intercomparison paper \citep{joos2013carbon}.

\subsubsection{Calibrating model parameters: the strength of the terrestrial carbon sink}

The IPCC experiments revealed a general tendency of intermediate complexity ESMs to understate the magnitude of the terrestrial carbon sink (the anthropogenic CO$_2$ taken up by vegetation on land). The major uncertainty in the terrestrial sink relates to CO$_2$ fertilisation. Experimental evidence almost without exception shows a stimulation of leaf photosynthesis when plants are exposed to elevated CO$_2$ \citep{korner2006plant}. In addition to this direct affect on photosynthesis, the short time-scale physiological effect of reduced stomatal opening increases water-use efficiency and additionally increases the efficiency of photosynthesis \citep{field1995stomatal}. However, the strength of the fertilisation effect is poorly quantified, especially under natural conditions. Some studies have failed to detect a measurable effect in nature, while others suggest that any effects may be short term, as CO$_2$ is only one of a number of potentially limiting factors on plant growth \citep{korner2006plant}. 

We addressed this calibration problem in \cite{holden2013model}. Using output from a 671-member ensemble of transient GENIE-1 simulations derived from the EFPC parameter sets we built an emulator of the change in atmospheric CO$_2$ concentration change since the preindustrial period. We then applied this emulator to sample the 28-dimensional input parameter space. A Bayesian calibration  suggests that the increase in gross primary productivity (GPP) in response to a doubling of CO$_2$ from preindustrial values is very likely (90\% confidence) to exceed 20\%, with a most likely value of 40-60\%. 

\subsubsection{Model understanding: what determines the spatial distribution of dissolved carbon in the ocean?}

In \cite{holden2013controls} we applied the EFPC ensemble to a transient experiment over the recent industrial era (1858 to 2008 AD). The temporal evolution of atmospheric CO$_2$ and its isotopic composition are known from observational data, and 
these simulated quantities were made to follow the observations through a relaxation term. 
The objective of the experiment was to better understand the mechanisms by which the anthropogenic CO$_2$ emissions are taken up by the ocean. 

To achieve this, we analysed the change in distributions of ocean concentrations of dissolved inorganic carbon (DIC) and its stable isotope $\delta^{13}C$, considering 
two-dimensional latitudinal-vertical transects through the Atlantic and Pacific. These two transects were combined into a single vector for each simulation (to ensure inter-basin effects were consistently represented), and these vectors were combined into an ensemble matrix. Singular vector decomposition was applied to the DIC and $\delta^{13}C$ matrices in order to extract the dominant modes of their spatial variability across the ensemble. Emulators of the component scores elicited further understanding of these modes by identifying which model parameters were driving 
each mode of variability. 

This, together with physical interpretation of the spatial patterns of each mode, enabled us to 
identify the principal processes driving them, on the assumption that the dominant parameters governing uncertainty in the response of each mode could be identified with the most important parameterised processes controlling the respective modes.
We showed that the 
main processes governing the uptake of anthropogenic CO$_2$ and $\delta^{13}C$ are quite distinct:
an important conclusion because observations of the isotopic composition are used to infer rates of 
ocean CO$_2$ 
uptake. Uncertainty in anthropogenic $\delta^{13}C$ uptake is 
dominated by air-sea gas exchange, which explains 63\% of modelled variance. This mode of variability is largely absent from the ensemble variability in CO$_2$ uptake, which is  instead driven by uncertainties in mixing rates between the surface and deep oceans. 

 \subsubsection{Coupling applications: coupling climate models and climate change impact models}

The evaluation of climate impacts requires coupling climate models, impact models and economic models together within an ``Integrated Assessment Model'' (IAM) framework. In such couplings, climate data (e.g. regional temperature, precipitation) are passed to the IAM for computation of climate impact functions, and the IAM passes back anthropogenic forcing (such as $CO_2$ emissions or land use change). Computational demands mean that it is generally infeasible to couple complex climate models into IAMs. Various approaches are taken to address this, using either simplified models or statistical representations of more complex models. Recently,  effort has focussed on the use of 
emulators of climate models as surrogates for the simulator in these coupling applications. 

Economic models provide projections of CO$_2$ emissions. They typically convert 
emissions into concentrations through the use of simple ``box-models'', describing rates of carbon transfer between the atmospheric, terrestrial and oceanic reservoirs. We have recently applied the EFPC parameter set to build an emulator of the
GENIE-1 carbon cycle model for incorporation into integrated assessment models \citep{foley2015modelling}. An 86-member subset of the EFPC parameter set was used to generate an ensemble of future climate-carbon cycle experiments,  with future emissions prescribed as modified Chebyshev polynomials.

The emulation approach followed the ``1-step'' dimensionally-reduced emulation methodology of \cite{holden2015emulation}, emulating a singular value decomposition of the ensemble outputs. Emulators of the first four component scores were derived as functions of the 28 model parameters and the 6 concentration profile coefficients. The emulator outputs are, unsurprisingly, dominated by the Chebyshev coefficients. However, uncertainty for a given forcing scenario is generated through emulator dependencies on GENIE-1 parameters. The resulting carbon cycle emulator has been coupled into an integrated assessment framework that also includes a macroeconometric model of the global economy E3MG \citep{mercure2014dynamics}, an agent-based model of technology substitution dynamics FTT-power \citep{mercure2012ftt} and a spatiotemporally resolved emulator of the climate system \citep{holden2014plasim}. We have applied the framework to assess the impact on the climate of emissions reduction policies in the electricity sector \citep{mercure2014dynamics}, addressing the cascade of uncertainty through the coupled system.

\section{Future applications}
 
It may never be possible to apply statistical approaches to robustly calibrate a truly state-of-the-art climate simulator. They are defined by the limits of available computing power, and consequently very few simulations are possible with these models. This begs the important question of how far could one go with simulator complexity and still be able apply these methods.
We have demonstrated the application of emulator-informed ABC to generate a 471-member ensemble of a model that takes $\sim10$ days to perform each simulation. The computational constraints ultimately determined the number of parameters we could vary; a rule of thumb dictates that we use  a minimum of 10 ensemble members for each varied active input \citep{Loeppky2009}. It is worth noting that a useful ensemble varying only, say, 5 parameters would need $\sim 50$ simulations, and could have been achieved for a 10-fold slower model.

The improvements in methodology demonstrated in Section \ref{section:toy}, suggest efficiencies that should significantly extend the applicability of the approach. The use of GP emulation generally allows a better statistical model than linear regression, and therefore would be expected to improve the success rate of the emulator filtering. This will certainly be the case when a parametric mean function is used and the GP is applied only to emulate the residual. The uncertainty estimates provided by the GP should also improve the success rate of the emulator filtering, for instance, by only accepting parameters for which there is a high  probability of plausibility. Furthermore, a significant improvement arises from the use of a sequential design process, which was shown to yield a 3-fold increase in efficiency in our example. For more complex simulators, we will want to make use of parallel computation. The sequential approach then changes from adding one design point at a time, to adding $d$, where $d$ is the number of available cores. Finding the $d$ optimal points that minimize the expected entropy is difficult, and is an area of active research, but even suboptimal designs can give significant improvements over the default space-fillings designs. For stochastic simulators, many of the same techniques can be applied. The likelihood function now needs to be estimated, significantly increasing the difficulty, but progress is being made in this direction \citep{Oakley2014, Meeds2014, Wilkinson2014, Gutmann2015}.
  
These improvements in efficiency should render application to ``previous-generation''  ESMs such as HadCM3\footnote {HadCM3 performs more than 10 years per CPU day on eight nodes of a linux cluster.} tractable on multi-node computing clusters, certainly so on distributed computing systems such as climateprediction.net, which last year facilitated more than 7,500 years of climate modelling on the personal computers of the general public.

 \bibliographystyle{abbrvnat}

{\small \bibliography{ABC_climate}}

\clearpage

 \end{document}